# Synthesis of Omnidirectional Path Loss Model Based on Directional Model and Multi-Elliptical Geometry

Jarosław Wojtuń[1], Cezary Ziółkowski[1], Jan M. Kelner[1], Tomas Mikulasek[2], Radek Zavorka[2], Jiri Blumenstein[2], Petr Horký[2], Aleš Prokeš[2], Aniruddha Chandra[3], Rajeev Shukla[3], Anirban Ghosh[4]

[1] Institute of Communications Systems, Faculty of Electronics, Military University of Technology, Warsaw, Poland, jaroslaw.wojtun@wat.edu.pl
[2] Department of Radio Electronics, Brno University of Technology, Brno, Czech Republic
[3] 'ECE Department, National Institute of Technology, Durgapur, India
[4] ECE Department, SRM University AP, 522240 India

*Abstract*—Millimeter wave (mmWave) technology offers high throughput but has a limited radio range, necessitating the use of directional antennas or beamforming systems such as massive MIMO. Path loss (PL) models using narrow-beam antennas are known as directional models, while those using omnidirectional antennas are referred to as omnidirectional models. To standardize the analysis, omnidirectional PL models for mmWave ranges have been introduced, including TR 38.901 by 3GPP, which is based on measurements from directional antennas. However, synthesizing these measurements can be complex and time-consuming. This study proposes a numerical approach to derive an omnidirectional model from directional data using multi-elliptical geometry. We assessed the effectiveness of this method against existing PL models for mmWaves that are available in the literature.

*Index Terms*—propagation, path loss (PL) model, directional PL model, omnidirectional PL model, antenna pattern synthesis, multi-elliptical propagation model, millimeter waves.

## I. Introduction

The development of mobile networks and the higher and higher requirements imposed on them force the use of increasingly larger radio resources [1]. Technological neutrality is conducive to this, as it allows for the replacement of older generations of systems (e.g., third generation (3G), analog television) by newer ones and the use of their spectral resources. On the other hand, the possibility of allocating new bands in the sub-6 GHz range (i.e., frequency range (FR)1) for mobile networks is becoming increasingly difficult. Hence, in fifth-generation (5G) and beyond networks, the potential of using millimeter waves (mmWaves) (i.e., FR2, 24–90 GHz, formally 30–300 GHz) has been noticed [2]. Wide FR2 bands allow for high throughputs, which ensures the possibility of creating efficient ultra-density networks (UDNs) [3]. At the same time, higher frequency ranges are characterized by limited range resulting from greater attenuation in free space and sensitivity to atmospheric conditions. This also has its advantages, because the limited radio range allows for redundant and compatible use of the same frequency bands in a small area. The potential of terahertz waves (formally 0.3–3 THz) is also planned to be used in sixth-generation (6G) networks [4]. Millimeter and terahertz waves increasingly resemble optical waves in their nature because they are characterized by greater directionality than lower-frequency radio waves. This feature requires the use of directional antennas (e.g., horn antennas) or multi-antenna systems with beamforming (e.g., massive or ultra-massive multiple-input-multiple-output (MIMO) systems) [5]. Currently, the use of higher ranges of centimeter waves (i.e., FR3, 7–24 GHz) is also being considered in 6G [6].

Most path loss (PL) models available in the literature refer to omnidirectional antenna systems, which results from their common use for lower radio frequency ranges. Currently, classical PL models are sometimes called omnidirectional, by analogy to the used antenna type. In the mmWave ranges, ensuring communication over longer distances requires using narrow-beam directional antennas or antenna systems with beamforming. However, using these antenna types disturbs the determination of classical PL models. Therefore, such models are called directional PL models. They are usually obtained by performing measurements using transmitting and receiving narrow-beam antenna systems directed at each other (i.e., under beam alignment conditions). To compare the changes in attenuation as a function of frequency, omnidirectional PL models for mmWaves have also been introduced in the literature [7–9]. These models are created by performing a procedure of synthesis of measurement results using narrow-beam antenna systems. For this purpose, measurements at each site must be made for different antenna positions in the azimuth and elevation planes on the transmitting and receiving sides, so as to provide a synthesized omnidirectional antenna pattern [10–11]. In practice, since mmWaves are strongly attenuated, synthesis is limited to a certain angle range relative to the direction between the transmitter (Tx) and receiver (Rx) [12].

In this paper, we propose a numerical approach to determining the omnidirectional PL model based on the directional model obtained from measurements. For this purpose, we use a multi-elliptical propagation model (MPM) [13–14], i.e., a geometric structure that allows modeling of the scattering occurring in the radio channel. The main advantage of this solution is the saving of time associated with performing measurements for the purpose of synthesizing the omnidirectional antenna patterns. The effectiveness of the

This research was funded in part by the National Science Center (NCN), Poland, grant no. 2021/43/I/ST7/03294 (MubaMilWave). For this purpose of Open Access, the author has applied a CC-BY public copyright license to any Author Accepted Manuscript (AAM) version arising from this submission.

developed method is assessed based on empirical directional and omnidirectional PL models for the selected frequency range determined by Prof. T. Rappaport's team from the NYU Wireless Research Centre. In our analysis, we are based on close-in free space reference distance (CI) PL models presented in [12].

The remainder of the paper is organized as follows. Sections II and III describe MPM and the numerical synthesis procedure, respectively. Exemplary results of the developed method for a selected frequency range are presented in Section IV. Finally, we conclude the paper.

## II. MULTI-ELLIPTICAL PROPAGATION MODEL

The methodology for transforming the propagation model, based on data obtained from measurements using a directional antenna system, into a model with an omnidirectional antenna system relies on the synthesis of the angular power distribution (PAS) around the receiving antenna. The multipath propagation environment is the cause of the occurrence of many components (replicas) of the same signal at the reception point, which differ in both power and delay time. It follows that the received signal is a superposition of components represented by a specific number $N$ of time clusters defined by a specific delay time relative to the propagation time of the wave in the direct direction between the transmitter and the receiver. In addition to the components that reach the receiver with a delay, there are also components resulting from local scattering, which occur in the vicinity of the receiving antenna. In the analyzed propagation model, we assume that the radiation/reception characteristics of the antenna power allow the analysis of propagation phenomena to be limited to the azimuth plane $\varphi$. Therefore, the angular power distribution $P(\varphi_R)$ of the received signal can be presented in the form:

$$P(\varphi_R) = \sum_{i=1}^{N} P_i(\varphi_R) + P_0(\varphi_R), \qquad (1)$$

where $P_i(\varphi_R)$ is the angular power distribution of all components that arrive at the receiver with delay $\tau_i$ and $P_0(\varphi_R)$ represents the PAS of all components that arise as a local scattering result.

The analysis of the two basic parts of (1) is carried out in a different way. In the MPM, to determine the sum of the components $P_i(\varphi_R)$, a geometric structure in the form of a set of confocal ellipses is used. This geometric structure is an extension of the model proposed in [15]. This set represents the statistical arrangement of scattering elements in space, which determines the mapping of propagation paths between the transmitter and receiver in a multipath propagation environment. The spatial configuration of this model, specifically the location of the foci of all ellipses, defines the coordinates of the Tx and Rx, which are separated by a distance $D$. The parameters of this model, i.e. the major axes $a_i$ and minor axes $b_i$ of individual ellipses, are closely related to the parameters that result from the characteristics describing the transmission properties of the propagation environment. The channel impulse response (CIR) or power delay profile (PDP) is the basis for determining such parameters as the values of local extremes (powers representing individual clusters) and their corresponding delays $\tau_i$. The relationship between the parameters describing the geometric structure and the signal parameters is described by the dependencies

$$2a_i = r_{1i} + r_{2i} = D + c\tau_i, \qquad (2)$$

$$2b_i = \sqrt{c\tau_i(c\tau_i + 2D)}, \qquad (3)$$

where c is the propagation speed of the electromagnetic wave in the medium.

The angular power distributions $P_i(\varphi_R)$ for individual ellipses determine the trajectories of propagation paths within each of them, and the powers of individual components are determined based on the even division of power falling on each of the clusters.

In the case of signal components that arise as a local scattering result, we assume that the statistical properties of the signal reception angle $\varphi_R$ are described by the von Mises distribution

$$f(\varphi_R) = \frac{\exp(\gamma \cos\varphi_R)}{2\pi I_0(\gamma)}, \qquad (4)$$

where $I_0(\cdot)$ is the zero-order modified Bessel function and $\gamma$ is the distribution parameter describing the intensity of the local scattering. The value of the $\gamma$ parameter of this distribution is selected depending on the analyzed propagation scenario. This makes it possible to obtain an omnidirectional or "strongly" unimodal distribution.

Thus, the angular power distribution $P_0(\varphi_R)$ of the components that arise as a local scattering result is described by the relationship

$$P_0(\varphi_R) = P_0 \cdot f(\varphi_R), \qquad (5)$$

where $P_0$ is the extreme value of PDP, which occurs for delay $\tau = 0$.

The geometric structure of the MPM model is presented in Fig.1 [13].

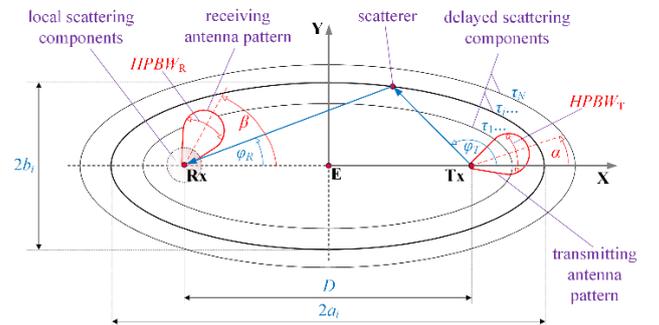

Fig. 1. Geometric structure of the MPM [13].

The geometric structure created by the set of ellipses is the basis for the correction of the angular power distribution obtained for the directional antenna system with respect to the omnidirectional antenna system. The essence of this correction consists of using the differences in the radiation/reception characteristics of the directional and omnidirectional antenna system and the assumed PAS. The synthesis of an omnidirectional path loss model is based on the formula defined the power $P$ of the received signal

$$P = \int_{-\pi}^{\pi} P(\varphi_R) d\varphi_R. \quad (6)$$

### III. NUMERICAL SYNTHESIS PROCEDURE

The proposed numerical procedure for the synthesis of an omnidirectional PL model is implemented according to the algorithm presented below. The directional PL model and parameters characterizing the transmitting and receiving antenna systems are the basic input data for this procedure.

---

**Algorithm 1** (*Numerical synthesis of omnidirectional PL model*)

1. Analysis of measurement scenario (i.a., frequency, $f$, distance range) and antenna parameters (i.e., half-power beamwidth, $HPBW$, gain, $G$) for the reference directional PL model, $PL_{\text{direct}}(D)$.
2. Selection of PDP resulting from measurements or use of PDP from a standard model, e.g., the 3GPP model [9] for the analyzed frequency.
3. Tx–Rx distance discretization $D = \{D_1, D_2, ..., D_j, ..., D_M\}$, where $j = 1,2, ..., M$, $D_j - D_{j-1} = \Delta D = $ const., Tx-Rx distance range $[D_1, D_M]$ results from measurement scenario.
4. For each analyzed Tx–Rx distance $D_j$, $j = 1,2, ..., M$:
   **repeat**
   5. Attenuation value determination of the reference PL directional for the analyzed Tx–Rx distance $D_j$, $PL_{\text{direct}}(D_j)$.
   6. Based on MPM, determination of PAS for directional antennas, $P_{\text{direct}}(\varphi_R)$.
   7. Based on PAS, $P_{\text{direct}}(\varphi_R)$, and Eq. (6), determination of received power for directional antennas, $P_{\text{direct}}$.
   8. Based on MPM, determination of PAS for omnidirectional antennas, $P_{\text{omni}}(\varphi_R)$.
   9. Based on PAS, $P_{\text{omni}}(\varphi_R)$, and Eq. (6), determination of the received power for omnidirectional antennas, $P_{\text{omni}}$.
   10. Determination of the attenuation for a synthesized omnidirectional antenna, $PL_{\text{omni}}(D_j)$ based on the following equation:

   $$PL_{\text{omni}}(D_j)(\text{dB}) = PL_{\text{direct}}(D_j) + 10 \cdot \log_{10}\left(\frac{P_{\text{omni}}}{P_{\text{direct}}}\right) \quad (7)$$

   **until** $D_j \leq D_M$
11. Determination of the omnidirectional PL model, $PL_{\text{omni}}(D)$, (e.g., CI PL model) based on the least squares method and discrete attenuation values $PL_{\text{omni}}(D_j)$ obtained for $D_j$.

---

The developed numerical procedure allows for fast estimation of an omnidirectional PL model based on an empirical directional PL model. In Section IV, we discuss the effectiveness of this approach based on empirical directional and omnidirectional PL models for two selected mmWave bands under line-of-sight (LOS) and non-LOS (NLOS) conditions.

### IV. EXEMPLARY RESULTS

To determine the relationship between PL and the Tx–Rx distance, we used the CI model [16]:

$$PL(D)(\text{dB}) = PL(D_0) + 10n\log_{10}\left(\frac{D}{D_0}\right) + X_\sigma, \quad (8)$$

where $n$ is the path loss exponent (PLE), which indicates the rate at which the path loss increases with distance, $D_0$ is the close-in reference distance determined from measurements close to the transmitter, $D$ is the Tx–Rx distance, and $X_\sigma$ is a zero-mean Gaussian random variable (in dB) with the standard deviation $\sigma$.

The effectiveness evaluation of the developed numerical synthesis method was performed based on omnidirectional and directional CI PL models (NYU) determined from measurements [12]. In this case, we selected models for two mmWave bands, 38 and 73 GHz. A detailed description of the measurement scenarios, used antenna systems, and obtained PL model parameters is presented in [12]. Unfortunately, there is no suitable PDP in [12]. Therefore, to generate the multi-elliptical scattering structure, we used the normal-delay profile for the urban macro (UMa) scenario and tapped delay line (TDL) models, i.e., TDL-B and TDL-D for LOS and NLOS conditions, respectively, from the 3GPP standard model [9]. We use a Sinc pattern model for shaping the directional antenna patterns in MPM [17].

Table I includes the most important parameters of these scenarios and the obtained PLEs for the synthesized omnidirectional PL models. The PL graphs for the individual scenarios are shown in Figs. 2–5 for 38 and 73 GHz under LOS and NLOS conditions, respectively.

TABLE I.    COMPARISION OF OMNIDIRECTIONAL PLEs

| Scenarios | Conditions | NYU models [12] | | Synthesized model based on MPM |
|---|---|---|---|---|
| | | $n_{\text{direct\_NYU}}$ | $n_{\text{omni\_NYU}}$ | $n_{\text{omni\_MPM}}$ |
| $f = $ 38 GHz, $HPBW_{Tx} = HPBW_{Rx} = $ 7.8°, $G_{Tx} = G_{Rx} = $ 25 dBi | LOS | 1.9 | 1.9 | 1.8 |
| | NLOS | 3.3 | 2.7 | 2.5 |
| $f = $ 73 GHz, $HPBW_{Tx} = HPBW_{Rx} = $ 7.0°, $G_{Tx} = G_{Rx} = $ 27 dBi | LOS | 2.3 | 2.0 | 2.1 |
| | NLOS | 4.7 | 3.4 | 3.6 |

We see that the obtained PLEs for empirically and numerically synthesized PL models are similar. Therefore, we can deduce that the proposed procedure allows for good estimation of omnidirectional PL models.

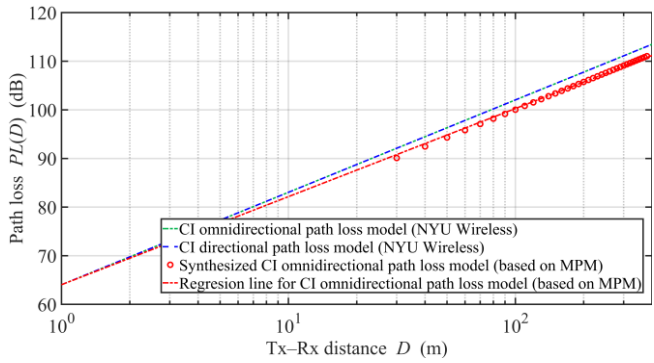

Fig. 2. CI PL models for 38 GHz under LOS conditions.

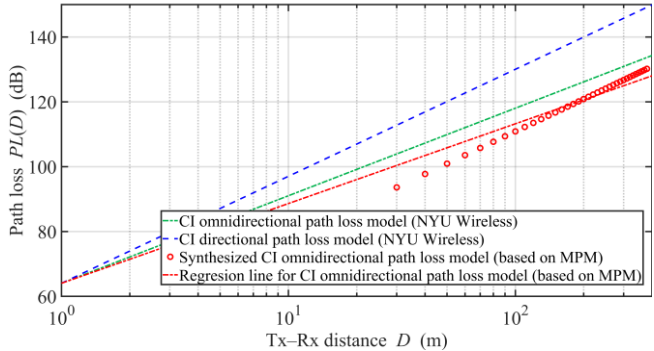

Fig. 3. CI PL models for 38 GHz under NLOS conditions.

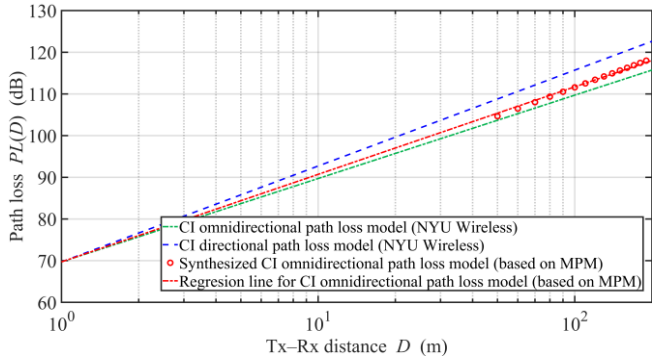

Fig. 4. CI PL models for 73 GHz under LOS conditions.

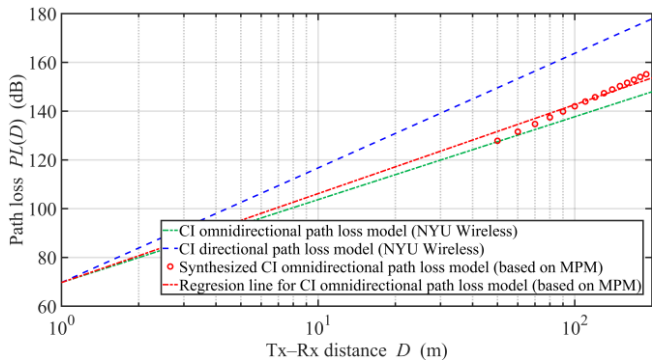

Fig. 5. CI PL models for 73 GHz under NLOS conditions.

We evaluate our model using the root mean squared error (RMSE) and mean absolute error (MAE) metrics [18]:

$$RMSE \text{ (dB)} = \sqrt{\frac{1}{M}\sum_{j=1}^{M}\left(PL_{\text{NYU}}(D_j) - PL_{\text{MPM}}(D_j)\right)^2}, \quad (9)$$

$$MAE \text{ (dB)} = \frac{1}{M}\sum_{j=1}^{M}\left|PL_{\text{NYU}}(D_j) - PL_{\text{MPM}}(D_j)\right|, \quad (10)$$

where $PL_{\text{NYU}}(D)$ and $PL_{\text{MPM}}(D)$ are omnidirectional CI PL models based on measurements (NYU) and proposed synthesis procedure.

These metrics allow for a quantitative assessment of the estimation accuracy of the NYU omnidirectional PL model by the model obtained from the developed synthesis procedure. The metrics are deviation measures between the CI PL models obtained from the empirical (NYU) and numerical (MPM) synthesis procedures. The RMSE and MAE values for different scenarios and conditions are shown in Table II. In this case, the distance was changed from 20 m to 200 m with step $\Delta D = 1$ m, i.e., $D_j = \{20, 21, \ldots, 200\}$ (m) and $M = 181$.

TABLE II. BASIC ERROR METRICS

| Scenario | Condition | MAE | RMSE |
|---|---|---|---|
| $f = 38$ GHz | LOS | 1.66 dB | 1.70 dB |
| | NLOS | 4.51 dB | 4.61 dB |
| $f = 73$ GHz | LOS | 1.87 dB | 1.92 dB |
| | NLOS | 4.64 dB | 4.75 dB |

In summary, the average estimation errors are about 1.7–1.9 and 4.5–4.8 dB for LOS and NLOS conditions, respectively. In each of the LOS/NLOS condition groups, the results obtained for both analyzed error measures and mmWave bands show a high degree of convergence. This proves the correctness of the presented synthesis method.

## V. CONCLUSIONS

In this paper, we presented a novel and original numerical synthesis procedure of omnidirectional PL models. The proposed tool and methodology enable fast estimation of an omnidirectional model based on an empirical directional one. This solution allows for the comparison of propagation conditions for different frequency ranges and measurement scenarios. Due to high attenuation, the implementation of measurements in the mmWave range prevents the widespread use of omnidirectional antennas and forces the use of directional antennas. Due to the use of narrow beam antenna systems, this is of great importance especially for mmWave bands. On the other hand, the developed solution allows for the verification of measurement results that were carried out using channel sounders or test-beds equipped with antenna systems with different parameters and patterns.

To obtain omnidirectional PL models, in the traditional approach, numerous measurements are performed with directional antennas, and then the omnidirectional antenna patterns are synthesized. This process is time-consuming and burdened with errors related to the overlap of beams from neighboring directions. The use of a numerical procedure simplifies the synthesis process, and the estimation errors are at an acceptable level.

In the near future, in the proposed procedure, the authors plan to replace the two-dimensional (2D) scattering structure (i.e., MPM) with a three-dimensional (3D) multi-ellipsoidal propagation model [19]. We hope that this could improve the accuracy of the PL estimation. On the other hand, we want to evaluate the effectiveness of the developed procedure in a wider scope by using the literature results for different mmWave bands.

ACKNOWLEDGMENT

This work was co-funded by the Czech Science Foundation under grant no. 23-04304L, the National Science Centre, Poland, under the OPUS call in the Weave program, under research project no. 2021/43/I/ST7/03294 acronym 'MubaMilWave' and by the Military University of Technology under grant no. UGB/22-748/2024/WAT, and chip-to-startup (C2S) program no. EE-9/2/2021-R&D-E sponsored by MeitY, Government of India.